# AN LSTM-BASED CHORD PROGRESSION GENERATION SYSTEM USING CHROMA HISTOGRAM REPRESENTATIONS

**Jack Hardwick**
Department of Musicology
University of Oslo
jackeh@uio.no

## 1. ABSTRACT

This paper proposes a system for chord generation to monophonic symbolic melodies using an LSTM-based model trained on chroma histogram representations of chords. Chroma representations promise more harmonically rich generation than chord label-based approaches, whilst maintaining a small number of dimensions in the dataset. This system is shown to be suitable for limited real-time use. While it does not meet the state-of-the-art for coherent long-term generation, it does show diatonic generation with cadential chord relationships. The need for further study into chroma histograms as an extracted feature in chord generation tasks is highlighted.

## 2. INTRODUCTION

Machine learning-based systems for chord generation have been employed in various contexts including as composition aids [1] and as components for interactive music systems [2]. Most implementations frame the task as a classification problem by predicting a suitable chord by way of discrete chords labels, as in [1], [3]. Frequently, systems can predict from a set of 12 major and 12 minor chords, although others have expanded chord vocabularies to include suspensions, inversions, and extended chords [2]. These systems have been implemented effectively as chord suggestions tools for composers and novice musicians.

For applications requiring more complex harmonic generation, such as advanced interactive music systems, a set of 24 chords is harmonically limiting. Such systems should be capable of generating diverse harmonies and therefore require at significantly larger chord vocabulary. Many-hot encoding has been used extensively for symbolic music generation tasks, such as in [4], leading to encouraging results.

Building on this work, this paper proposes the use of chroma histogram representations as extracted features for chord generation, thereby framing this as a regression problem. This is in pursuit of a chord representation more suitable for real-time applications. In turn, I propose that this representation can expand the harmonic possibilities afforded by chord generation models.

Chroma histograms provide a compressed and low-dimensional representation of harmonic information, in turn making models less computationally intensive to train and use in performance. With chord predictions taking approximately 80ms on average, this work also demonstrates the potential of such models for real-time use. The results of this work, while limited in the long-term musical coherence of the chords generated, will constitute a basis for further exploration into chroma representations for chord generation in real-time scenarios.

## 3. RELATED WORK

Previous researchers have developed several machine-learning based systems for chord suggestion and prediction. Primarily these systems generate chords through one-hot encoding or classification approaches, such that the output represents a single chord or set of related chords. In [5] Paiement et al. present four approaches for chord representation in this manner. The 'naïve' representation considers every possible chord as a separate class, the 'root' representation represents chords as only their root with chord type (major/minor) information being assumed from the key, the 'mM7' representation which groups similar chords thereby reducing vocabulary size, and a combination of the 'root' and 'mM7' approaches yielding a vocabulary of 36 chords.

In [3] which makes use of a bi-directional LSTM (BLSTM) architecture for chord generation, a chord is predicted from a set of 24 representing all standard Western major and minor chords. Lim, Rhyu and Lee implement a BLSTM architecture to improve performance in generating coherent chord sequences. They hypothesise that as chord order is an important component of coherent progressions, we must consider reverse chord order to also be important. However, their BLSTM model iterates through a given melody in both forwards and backwards. It is therefore unsuitable for real-time applications, in which the melody can only be iterated through in the forward direction.

A related system is employed in [1] for ChordRipple, a chord suggestion system using natural language processing techniques. This system maps chords in latent space by their semantic similarity, such that chords that are used in similar contexts or fulfil similar harmonic roles are places closely in the latent space. Huang, Duvenaud, and Gajos' model is therefore able to suggest chords that fulfil a similar semantic function i.e., resolution to a tonic, while also exploiting a wider range of chords than the 24 allowed by Lim et al. Specifically, their system digests chords symbols into attributes such as root note, type, extensions, inversions, and other aspects. As with my system, their system is intended to capture a wider range of harmonic possibilities. However, the mapping of chords within the latent space is a different approach to mine, which instead learns the semantic relationships between chords through their sequential nature in the training dataset.

Others have also implemented machine learning-based systems for generating accompaniment in real time. In [2], Garoufis et al. present an LSTM-based interactive system which employs a vocabulary of 121 discrete chord labels. The system selects several candidate chords for the performer to choose between. The system is controlled with an Xbox Kinect camera, with the user adopting the hand posture of a guitarist. Vertical movement in the right hand indicates strumming to trigger chords, while the position of the left hand along a virtual fretboard determines which of the possible chords is chosen. Like my system, Garoufis et al.'s is iterative as the chord progression held in memory is updated with each new sounding chord, which in turn updates the chord selection presented to the user.

Similarly, [6] presents SongDriver, a model for real-time accompaniment generation which aims to minimise logical latency in predictions, or the time taken to generate new accompaniment to a melody and reduce error propagation induced by improper predictions. SongDriver uses the Transformer model architecture to arrange chords in real-time for the melody and cache chords for future use, followed by a conditional random field (CRF) model to which arranges the cached chords into a multi-part accompaniment.

There has been limited previous study into the use of chroma histograms for chord recognition, prediction, or suggestion. In early work on the topic, Papadopoulis and Peeters propose a method of chord recognition from raw audio using extracted chroma histograms and Hidden Markov Models in [7]. Their method employs normalised chroma histograms to determine the probability of a given histogram corresponding to one of 24 major or minor chords. Similarly, Cho and Bello in [8] present an exhaustive evaluation of the most-utilised methods for chord recognition using chroma features. Like many, the systems they evaluate are designed to recognise the most common 24 chords, therefore are limited in their applicability to the design of my system.

## 4. SYSTEM DESIGN

Chords are extracted from the dataset of MIDI files through four processing stages: 4.1) key signature extraction/prediction; 4.2) melody instrument selection; 4.3) chord chroma histogram calculation, and 4.4) similar adjacent chord removal.

### 4.1 Key Signature Extraction

For chord generation datasets, it is common to first align all MIDI files to C major, as in [3], [4], [6]. This ensures that all chords are represented by their relationship to the tonic. I extract key data from MIDI files directly when available. Files in minor keys are interpreted as their relative major i.e., A minor is interpreted as C major.

To extract key information from MIDI files without key signature metadata, I train a Support Vector Machine classification model for key signature recognition. The input is normalised chroma histograms of MIDI files in the dataset represented by 1D vectors of length 12, while the output is an integer in range 0-11 representing the major keys from C major.

The dataset for this model is discussed in Section 5.1 below. The dataset is initially unbalanced. G, D, and A major are heavily over-represented among major keys, and major examples significantly outweigh minor examples by 5.89:1. Minor examples are duplicated 6 times, with the addition of random noise to each, to balance the number of major and minor examples. All examples are then transposed by a random number of semitones to negate the bias towards popular keys. This increases the dataset size from 5580 to 9626 examples, of which 4772 are major and 4854 are minor. Minor examples are then transposed to their relative major by adjusting their key labels.

The model is trained on this balanced dataset after PCA dimensionality reduction. The hyperparameters used are chosen through grid search cross-validation: RBF kernel, *C=0.5*, *degree=1*, and *gamma=1*. The balanced class weight is used due to the prior balancing stage. The model achieves an accuracy of 83.9% when trained on the major-transposed dataset, as opposed to 68.5% with 24 keys. As a result, the 12-key implementation is used.

### 4.2 Melody Instrument Selection

As the Lakh dataset is not consistent in the amount, program or name of the MIDI tracks used in each file, I algorithmically select an appropriate melody track from each file. The instrument names of all non-drum tracks are first checked for one of the following: 'voice', 'vocal', 'vox', 'sing', and 'melody'. If one or more are present, the melody instrument is selected from among these candidates.

If none of the above keywords are present, the proportion of overlap among notes within each non-drum instrument is calculated as a proportion in range 0 (no overlap) to 1 (full overlap), as a metric of the level of polyphony in the instrument. Instruments which have an overlap proportion below a threshold are selected as melody instrument candidates. Instruments with a lower proportion of overlap are more monophonic, and hence more likely to also be melodic. This method is shown in Figure 1.

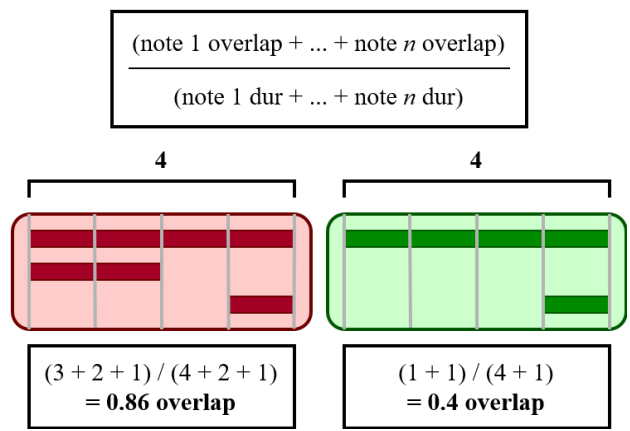

Figure 1: The formula and an example for calculating instrument overlap.

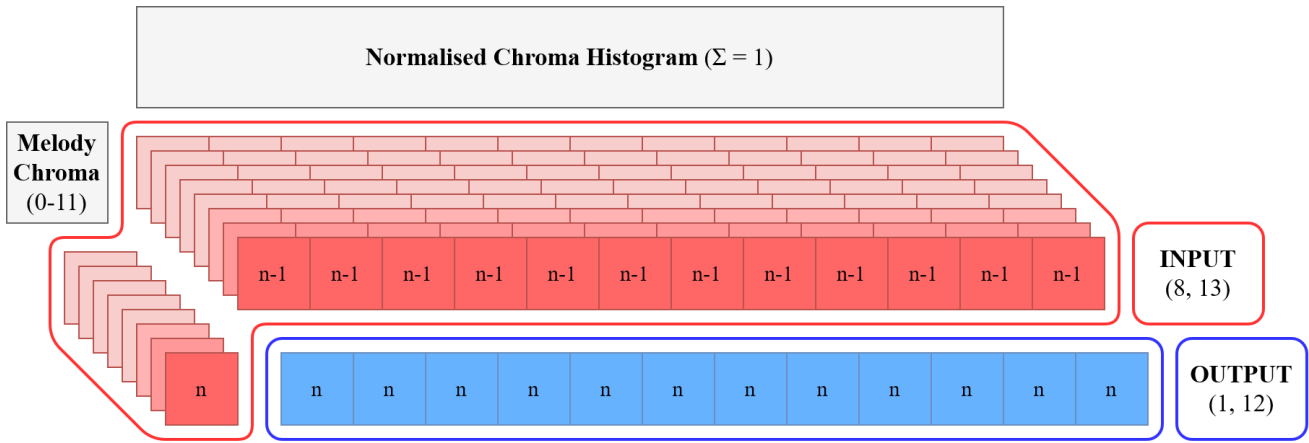

Figure 3: The input & output structure of the chord generation model.

From the list of melody instrument candidates, the instrument with the highest mean MIDI note number is used as the melody instrument, making basslines unlikely be selected as the melody instrument. This approach is adapted from [9] in which a similar method is used to distinguish between left- and right-hand piano parts in MIDI files.

### 4.3 Chord Chroma Histogram Calculation

For each melody note, a normalised chroma histogram is calculated from all overlapping notes in other non-drum tracks. This provides a compressed representation of the harmony which accompanies each note in the melody instrument. The melody instrument and the chroma histograms are aligned to C major using the key signature extracted either from the MIDI file or by the key signature classifier. This ensures the chroma pitch of the melody note and the chroma histogram describe their respective harmonic relationships to the tonic. Concurrent chords in the MIDI files are stored adjacently in the dataset to preserve chord order. The size of the dataset at this stage is 4,320,697 histogram and melody chroma pairs.

### 4.4 Similar Adjacent Chord Removal

Some adjacent chords in the dataset are removed to reduce the frequency of similar chroma histograms appearing sequentially. Garoufis et al. in [2] identify sequentially repeated chord information in datasets for music generation as a common barrier to diverse music generation, and this bore out in initial versions of my chord generation model.

Removing adjacent similar chords drastically improves the diversity of the chords generated in my model. Adjacent chroma histograms which are identical or have 4 or more non-zero histogram values are within a threshold of 0.1 of each other are removed from the dataset. This reduces the size of the dataset to 3,773,148 examples, a decrease of 8.7%.

### 4.5 Chord Generation Model

The chord generation model is a stacked LSTM architecture with an input layer, 3 LSTM layers with 50% dropout to reduce overfitting, and a dense output layer. LSTM layers are used to learn the long-term sequences and dependencies in the dataset, which are necessary for generating coherent chord progressions. I train the final version of the model using a sequence length of 8, a batch size of 64, and a learning rate of 0.0001. The architecture of the model is shown in Figure 2.

The input of the model is a 2D vector of shape (8, 13) representing the input sequence length and number of features respectively. The 13 features consist of chroma values for the current melody note $n$ and previous melody notes $n-1$ to $n-7$, and the chroma histogram for the chords $n-1$ to $n-8$. The output of the model is a 1D vector of length 12, which represents chord $n$, the normalised key-quantised chroma histogram to accompany the current melody

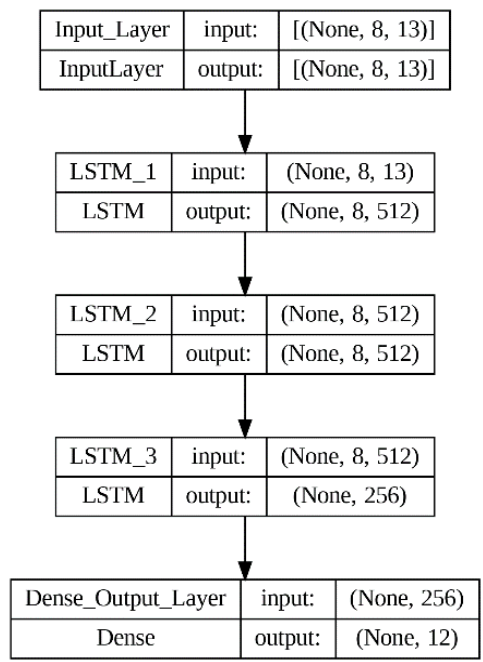

Figure 2: The architecture of the main chord generation model.

note. This input and output structure is shown in Figure 3.

The model utilises the Adam optimiser. The Mean Squared Logarithmic Error loss function does not penalise outliers in the dataset, and therefore diverse generation, as heavily as the standard Mean Squared Error.

### 4.6 Model Implementation & Chord Voicing

Due to time constraints, I do not implement the model for real-time use. Instead, the final system can iterate through a pre-composed melodic MIDI file, predict and voice chords, and write the output to another MIDI file. To voice chords, chroma which do not meet a threshold of 0.14 in the predicted histogram are disregarded. The remaining chroma with the largest value is the root of the chord and is voiced in octaves in the range of C2 – B3. The remaining chroma which exceed the threshold are voiced in the octave of C3-B3. All chords are transposed based on the supplied tonic of the input melody. An example of voicing a chord is shown in Figure 4.

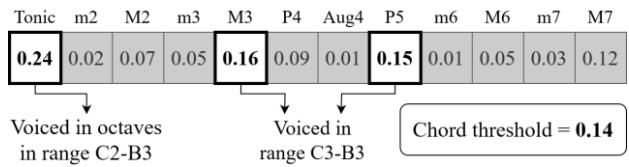

Figure 4: An example of voicing a tonic major chord.

## 5. DATASET

### 5.1 Key Signature Classification Dataset

The key classification model is trained using primarily MIDI files from the cleaned Lakh dataset [10] which contain valid key information. MIDI files containing C major at 0 seconds are excluded as these are often incorrect. To improve the robustness of the model across different musical styles, this dataset is supplemented with the POP909 [11] and Nottingham [12] MIDI datasets. The POP909 dataset consists of 909 songs of Chinese popular music. The Nottingham dataset consists of 1034 British and American folk songs.

### 5.2 Chord Generator Dataset

The chord prediction model is trained on chords extracted from the cleaned subset of the Lakh MIDI dataset, which are processed as discussed in Sections 3.1-3 above. I chose this dataset for its balance between Western musical convention and diversity of harmony. The dataset consists of over 17,000 MIDI files covering 2,200 artists. It consists of mostly conventional Western popular harmony from popular music. However, it also contains other genres including classical music and jazz, therefore the model can learn from a more diverse range of harmony than only standard popular progressions.

## 6. SYSTEM IMPLEMENTATION

The key signature classifier model is implemented using SciKitLearn. I extract chroma histograms from the dataset using PrettyMIDI. Key signature metadata from the Lakh and Nottingham datasets is extracted using PrettyMIDI. Key signatures for the POP909 dataset are extracted from the included text files.

The system is built in Python across nine Jupyter Notebooks: five are used in dataset preparation and training of the key signature classifier, two are used for processing the main dataset as discussed in Sections 4.1-4.4, and another is used for training the main LSTM model. The last notebook is used to apply the trained model to pre-composed MIDI files.

Processing and training were executed both locally and remotely. Processing of the datasets for the key classifier and the chord generator, and the training of the key classifier were all performed locally. Training of the chord generator was performed remotely using Google Colab Pro for its access to GPUs to improve training times, and to allow execution while the host PC is offline.

## 7. RESULTS & DISCUSSION

I evaluate my system through both qualitative and quantitative methods. Qualitative evaluation occurred as part of the design and training process, while quantitative evaluation consisted of a small user study.

### 7.1 Qualitative Evaluation of System Design

I engaged in qualitative evaluation of the system throughout the development process. I trained 3 versions of the model using sequence lengths of (4), (8), and (16) respectively. I selected the (8) model to be used for the generation evaluated in the user study below because it demonstrates the best balance between long-term coherence and minimising error propagation. The (4) model generates some recognizable chords in isolation but displays little long-term coherence. Meanwhile, the (16) model suffers from error propagation as discussed in [6] in Section 3 above. Prominent non-diatonic chroma pitches are fed back into the model, resulting in more non-diatonic notes being selected, causing the generation to slowly drift from the specified major key.

The (8) model demonstrates occasional music grammar. Its generation is largely diatonic with a strong preference towards the most common chroma values in the dataset. The generated chords are regularly recognizable in isolation. However, while the model occasionally generates cadential chord transitions (see listening excerpt 1A), these are generally constrained to pairs of adjacent chords, and rarely extend to sequences of three or more chords.

This shortcoming is likely due to the structure of the dataset. By generating a chroma histogram per melody note rather than per bar, the dataset does not contain metrical information therefore relationships such as 1 chord per bar are not maintained. Instead, chord order can only be learned sequentially by adjacent histograms in the dataset.

### 7.2 Qualitative Evaluation by User Study

In a similar fashion to [3], I recruited 5 participants completed a survey evaluating the chords generated by the system to 3 varied melodies, in comparison to the original chords. While Lim et al. compare the generation from their BLSTM model to other Hidden Markov Model-based systems, I only compare my system against the pre-composed chords. The melodies are intended to be unknown to the listener to not bias their opinions between the composed and generated chords. The respondents were all acquaintances of the author.

4 of the 5 respondents have extensive musical training.

Participants scored two accompaniments A and B to each of 3 melodies on a 1-7 Likert scale to the prompt 'the chords felt appropriate to accompany the melody'.

Participants were then asked to choose either accompaniment A or B for: a) which they found more engaging to listen to, and b) which they preferred. Accompaniments A and B were randomised between melodies such that the generated chords alternated between Accompaniment A and B. The listening examples can be found in the included project files. As expected, participants showed a strong preference towards the composed rather than generated chords for all three melodies. The results are summarised in Table 1.

Unexpectedly, Melody 1 performed worst between the melodies in the engagement and overall preference metrics. Melody 1 contains the highest rate of musical cadences among the generated accompaniments; hence I expected this melody to perform best overall between the melodies.

### 8. CONCLUSION

This paper proposes an LSTM-based model for chord generation to symbolic melody using chroma histogram representations. While the results of this work are not state-of-the-art regarding long-term coherence in chord generation, they demonstrate that chroma histogram representations merit further study for chord generation tasks. The system generates largely diatonic harmony to a melody and has evidently learnt some basic guidelines of Western popular harmony.

However, the lack of musical coherence in the output is currently a significant barrier to usability. Further work should focus first on the processing of the dataset. Specifically, calculating chroma histograms per bar, as discussed in Section 7.1, should allow the model to learn chord transitions more accurately, as this models the common structure of chords in the dataset more accurately. Melody chroma in this case could be extracted by selecting the most dominant chroma pitch in the melody instrument during the given bar.

Relatedly, extracting chroma histograms from smaller, more specialised datasets may allow the model to learn harmonic rules from certain styles or composers. For example, training on the JSB Chorales dataset of Bach chorales might teach the model to generate progressions in line with Bachian harmony. Alternatively, training on jazz standards may yield non-diatonic, jazz-like harmonic language.

The current implementation of the system creates 81ms of logical latency per chord prediction. A real-time version of the system would be workable for beatless music in which high latency can be accounted for. A real-time system using live MIDI pitch data would be straightforward to implement. A system involving pitch recognition on live recorded sound could also be implemented at the cost of additional logical latency.

Despite the limited results presented here, this work nonetheless proposes a novel use for chroma histogram representations for chord generation to symbolic melodies. Future study should focus on two key areas. Firstly, the construction of the chords dataset to improve the consistency of harmonically meaningful chord transitions from which the model can learn. Secondly, the real-time applicability of this model as a component for interactive music systems should be evaluated further.

|  | **Mean Appropriateness Score** (0-7) | | **More Engaging** (%) | | **Preferred** (%) | | **Knew melody** |
|---|---|---|---|---|---|---|---|
|  | Generated | Composed | Generated | Composed | Generated | Composed |  |
| **Melody 1** | 3.6 | 5.6 | 0% | 100% | 0% | 100% | 0% |
| **Melody 2** | 3.8 | 6.4 | 20% | 80% | 20% | 80% | 0% |
| **Melody 3** | 3.4 | 5.8 | 20% | 80% | 20% | 80% | 60% |
| **Mean** | 3.6 | 5.93 | 13.3% | 86.6% | 13.3% | 86.6% |  |

Table 1: Results from the user survey, showing preference for composed chords over generated chords for all 3 melodies.